\magnification=\magstep1
\font\mittel=cmbx10 scaled \magstep1
\font\gross=cmbx10 scaled \magstep2

\centerline{\gross Lie algebras with finite}

\centerline{\gross  dimensional polynomial centralizer}
\bigskip
\bigskip
\centerline{{\bf Giuseppe Gaeta}\footnote{*}{\sevenrm{Work supported by ``Fondazione CARIPLO per la ricerca scientifica'' under the project ``Teoria delle perturbazioni per sistemi con simmetria''}}}
\medskip
\centerline{Dipartimento di Matematica, Universit\'a di Milano}
\centerline{via Saldini 50, I--20133 Milano (Italy)}
\medskip
\centerline{Dipartimento di Fisica, Universit\'a di Roma}
\centerline{P.le A.~Moro 5, I--00185 Roma (Italy)}
\bigskip
\centerline{\bf Sebastian Walcher}
\medskip
\centerline{Lehrstuhl A f\"ur Mathematik, RWTH Aachen}
\centerline{D--52056 Aachen (Germany)}
\medskip
\centerline{Institut f\"ur Biomathematik und Biometrie, GSF-Forschungszentrum}
\centerline{Ingolst\"adter Landstr.~1, D--85764 Neuherberg (Germany)}
\bigskip
\bigskip
\noindent{\bf Abstract.} We give criteria for finite dimensionality or infinite dimensionality of the polynomial centralizer of the Lie algebra of a linear Lie group, in terms of invariants and relative invariants of the group. In the finite dimensional scenario some applications to normal forms and to certain equations with fundamental solutions are presented.
\bigskip
\noindent{\bf AMS Subject Classification (2000):} 34C14, 34A05, 22E05
\bigskip
\noindent{\mittel 0. Introduction.} 
\smallskip
\noindent We consider the following setting: Let $V$ be a finite dimensional vector space over ${\bf K}$ (standing for the real numbers ${\bf R}$ or the complex numbers ${\bf C}$), and let ${\cal M}\subseteq { g\ell}(V)$ be a Lie algebra. Considering the Lie algebra ${\cal P}(V)$ of all polynomial vector fields on $V$ (note that ${\cal M}$ may be identified with a subset of ${\cal P}(V)$), we are interested in the centralizer 
$${\cal C}({\cal M})=\{f\in {\cal P}(V):\,\left[ f, {\cal M}\right] =0\}.$$ 
In other words, ${\cal C}({\cal M})$ is the Lie algebra of infinitesimally ${\cal M}$-equivariant vector fields.

The purpose of this note is to present criteria for ${\cal C}({\cal M})$ to be finite dimensional or infinite dimensional over ${\bf K}$, and to discuss some aspects of the finite dimensional scenario.

There are two principal reasons to motivate this investigation. The first is derived from a connection to finiteness (or even triviality) of Poincar\'e-Dulac normal forms (see e.g. Bruno [2]), which may be imposed on a vector field by the existence of certain symmetries. A survey of the work in this field is contained in [6]. Second, according to a theorem of Lie [8], (non-autonomous) differential equations associated with finite dimensional Lie algebras of vector fields enjoy the ``fundamental solution property''; i.e., the general solution of such an equation can be expressed as a function of finitely many (sufficiently generic) particular solutions. This result of Lie was brought back to general attention, and studied in modern terms, by Winternitz [18, 19]; see also [7], Ch.~IX, Section 2, and the recent book by Cari\~nena, Grabowski and Marmo [3]. Many examples are known, the simplest ones being linear and (matrix) Riccati equations, and such differential equations are interesting for various reasons. Classifications of the equations satisfying certain nondegeneracy conditions are known; see Shnider and Winternitz [12, 13], or the connection between transitive algebras and Jordan pairs exhibited in [15]. Finite dimensional centralizers of linear Lie algebras, on the other hand, may be seen as a pathological class. The associated differential equations are interesting since they cannot be reduced via invariants of the symmetry algebra, whence a different approach has to be taken.
\bigskip
\noindent{\mittel 1. Characterization. }
\smallskip
\noindent By ${\bf K}\left[x_1,\ldots,x_n\right]$ we denote the algebra of polynomials in $n$ variables over ${\bf K}$. A vector polynomial $f=\left(\phi_1,\,\ldots,\,\phi_n\right)$ induces a derivation
$$X_f = \sum \phi_i\, \partial/\partial x_i$$
of  ${\bf K} \left[x_1,\ldots,x_n\right]$, and every derivation is of this type. We recall the rule 
$$X_fX_g-X_gX_f = X_{[f,\,g]}\quad{\rm with} \quad [f,\,g](x)=Dg(x)f(x)-Df(x)g(x).$$
Let ${\cal M}\subseteq { g\ell}(V)$ be a Lie algebra, and $\alpha$ a linear form on ${\cal M}$. We define vector spaces
$$I_{\alpha}({\cal M}):= \left\{\phi\in {\bf K}\left[x_1,\ldots,x_n\right]:\, X_B(\phi)=\alpha(B)\cdot \phi,\quad{\rm for\,\,all} \,\,B\in {\cal M}\right\},$$
and call these the spaces of {\it relative $\alpha$-invariants} of ${\cal M}$. In particular, 
$$I_{0}({\cal M})= \left\{\phi\in {\bf K}\left[x_1,\ldots,x_n\right]:\, X_B(\phi)=0 \quad{\rm for\,\,all} \,\,B\in {\cal M}\right\}$$
is the space of all polynomial {\it (infinitesimal) invariants} of ${\cal M}$. Note that $I_{0}({\cal M})$ is also an algebra, and that each $I_{\alpha}({\cal M})$ is canonically a module over this algebra. Regarding our topic, we recall a well-known fact:
\vfill\eject
\noindent{\bf (1.1) Proposition.}
\smallskip
\noindent If $I_{0}({\cal M})\not = {\bf K}$ then ${\cal C}({\cal M})$ has infinite dimension.
\smallskip
\noindent{\it Proof.} Let $E(x):=x$ for all $x$, and let $\phi$ be a nonconstant invariant. Then $\phi^m\cdot E \in {\cal C}({\cal M})$ for all $m$.

\rightline{$\diamond$}
\medskip
In particular, if ${\cal M}$ is the Lie algebra of a compact linear group then ${\cal C}({\cal M})$ is always infinite dimensional, since nontrivial invariants exist. More generally, for reductive groups it is known that ${\cal C}({\cal M})$ is a finitely generated module over $I_{0}({\cal M})$. (The papers by Sartori [11], and Schwarz [14] contain the relevant information on reductive groups; further references can also be found there.) Thus, for reductive groups the converse of Proposition 1.1 holds true.

\noindent The following is a partial converse in the general setting:
\medskip
\noindent{\bf (1.2) Proposition.} 
\smallskip
\noindent If ${\cal C}({\cal M})$ has infinite dimension then some $I_{\alpha}({\cal M})$ has infinite dimension.
\smallskip
\noindent{\it Proof.} Let ${\bf L}={\bf K} \left(x_1,\ldots,x_n\right)$ be the field of rational functions over ${\bf K}$.

\noindent (i) Let $\left(f_i\right)_{i\geq 1}$ be a family of elements of ${\cal C}({\cal M})$ that is linearly independent over ${\bf K}$. (Note that the degrees of the $f_i$ are unbounded.) Considered as a family of elements of ${\bf L}^n$, it is linearly dependent. We may assume that $(f_1,\ldots,f_s)$ is a maximal linearly independent subsystem. Then there are  polynomials $\mu_{ij}$, $\theta_{ij}$ (which may be chosen relatively prime) such that 
$$f_j= \sum_{i=1}^s {{\mu_{ij}}\over{\theta_{ij}}}\cdot f_i, \quad{\rm all}\,\,j>s.\leqno{(\dagger)}$$
Moreover, $X_B(\mu_{ij}/\theta_{ij})=0$ for all $i$, $j$ and all $B\in{\cal M}$, since
$$0=\left[B,\,f_j\right]=\sum_{i=1}^s X_B({{\mu_{ij}}\over{\theta_{ij}}})\cdot f_i$$
for all $j>s$ and all $B\in{\cal M}$.
 (The argument is the same as the one used in [17].)

\noindent The invariance of $\mu_{ij}/\theta_{ij}$ forces
$$X_B(\mu_{ij})\cdot\theta_{ij} = X_B(\theta_{ij})\cdot\mu_{ij}.$$
Unless $X_B(\mu_{ij})=X_B(\theta_{ij})=0$, unique prime factorization in ${\bf K}\left[x_1,\ldots,x_n\right]$ implies that every prime factor of $\mu_{ij}$ must divide $X_B(\mu_{ij})$ with according multiplicity, and the same holds, {\it mutatis mutandis}, for $\theta_{ij}$. Thus
there is a scalar $\alpha_{ij}(B)$ so that $X_B(\mu_{ij})=\alpha_{ij}(B)\cdot \mu_{ij}$ and $X_B(\theta_{ij})=\alpha_{ij}(B)\cdot \theta_{ij}$, for each $B$. Clearly, $\alpha_{ij}$ is a linear form on ${\cal M}$. This shows that $\theta_{ij}$, as well as $\mu_{ij}$, is a relative invariant of ${\cal M}$ with associated linear form $\alpha_{ij}$.

\noindent (ii) Using Cramer's rule for the equation $(\dagger)$ and a suitable $s\times s$ minor of the matrix with columns $f_1,\ldots,\,f_s$ over {\bf L}, one sees that there is a polynomial $\theta^*$ which is a multiple of all $\theta_{ij}$. Moreover, for every $j$ there is an $i$ so that $\theta_{ij}$ or $\mu_{ij}$ is not constant, since the $f_i$ are linearly independent over ${\bf K}$. If $I_0({\cal M})={\bf K}$ we have $\theta_{ij}\not={\rm const}$. Since $\theta^*$ has only finitely many prime factors, there is a nonconstant $\theta$ such that $\theta_{ij}$ is a scalar multiple of $\theta$ for infinitely many $(i,\,j)$. Now $\theta$ is a relative invariant of ${\cal M}$ and we are done, because the degrees of the corresponding $\mu_{ij}$ are unbounded.

\rightline{$\diamond$}
\medskip
\noindent{\bf (1.3) Corollary.}
\smallskip
\noindent If ${\cal C}({\cal M})$ is infinite dimensional then ${\cal M}$ admits nontrivial rational invariants.

\rightline{$\diamond$}
\medskip
The question whether the converse of Prop.~1.1 is generally true remains open, but we can show this in a number of relevant cases.
\medskip
\noindent{\bf (1.4) Proposition.}
\smallskip
\noindent Assume that ${\cal C}({\cal M})$ is infinite dimensional, and that either $\left[{\cal M},\,{\cal M}\right] = {\cal M}$ or that ${\cal M}$ is solvable. Then $I_0({\cal M})$ is nontrivial.
\smallskip
\noindent{\it Proof.} (i) One has $\alpha([{\cal M},\,{\cal M}])=0$ whenever $I_{\alpha}({\cal M})\not= \{0\}$. To see this, note for $B$, $C\in{\cal M}$ and $\phi\in I_{\alpha}({\cal M})$:
$$X_{[B,C]}(\phi) = X_BX_C(\phi)-X_CX_B(\phi)=\alpha(C)X_B(\phi)-\alpha(B)X_C(\phi)=0.$$
Therefore, only $I_0({\cal M})$ is nonzero in case $\left[{\cal M},\,{\cal M}\right] = {\cal M}$.

\noindent (ii1) Now let ${\cal M}$ be solvable. It is harmless to assume ${\bf K}={\bf C}$. According to Lie's theorem on solvable algebras, ${\cal M}$ may be taken as an algebra of upper triangular matrices, thus
$$B=\pmatrix{\rho_1(B) & & * \cr
                       & \ddots & \cr
                     0 & & \rho_n(B)\cr}$$
for each $B\in {\cal M}$, with linear forms $\rho_i$.

\noindent (ii2) Let $\phi\not= 0$ be homogeneous of degree $d$, and suppose $\phi\in I_{\alpha}({\cal M})$. Then there are nonnegative integers $d_1,\ldots,d_n$ such that$$d_1+\ldots +d_n=d \quad{\rm and}\quad d_1\rho_1+\ldots +d_n\rho_n=\alpha,$$
and  
$$\psi:=x_1^{d_1}\cdots x_n^{d_n}\in I_{\alpha}({\cal M}).$$
To see this, note first that $X_B(x_i)\in {\bf K}x_i+\ldots {\bf K}x_n$ for all $i$ and all $B$, whence for $1\leq k_1<\ldots < k_r$:
$$\eqalign{
X_B(x_{k_1}^{d_{k_1}}\cdots x_{k_r}^{d_{k_r}})=& \left(d_{k_1}\rho_{k_1}(B)+\ldots +d_{k_r}\rho_{k_r}(B)\right)\cdot x_{k_1}^{d_{k_1}}\cdots x_{k_r}^{d_{k_r}}\cr
+ &\,\,{\rm terms}\,\,{\rm of}\,\,{\rm smaller}\,\,x_{k_1}\,{\rm -\,exponent}.\cr}$$
Now choose $l_1$ as the smallest index, and then $d_{l_1}$ as the highest exponent, among all the monomials occurring in $\phi$ with nonzero coefficient, and set $\psi:=x_{l_1}^{d_{l_1}}\cdots x_{l_r}^{d_{l_r}}$. Then the above considerations show $X_B(\psi)=\alpha(B)\cdot\psi$. Thus every nonzero homogeneous relative invariant yields a nonzero monomial relative invariant of the same degree, corresponding to the same linear form.

\noindent (ii3) Since $I_{\alpha}({\cal M})$ has infinite dimension, there is a sequence of monomials
$$\psi_l=x_1^{d_{1,l}}\cdots x_n^{d_{n,l}}\in I_{\alpha}({\cal M})$$
such that $e_l:=\sum_{i=1}^n d_{i,l}\to\infty$ as $l\to\infty$. Note that $\psi_l/\psi_k$ is a rational invariant of ${\cal M}$ for all $r$ and $s$. The argument in [16], part (b) of proof of Prop.~1.6, now applies verbatim and shows that some $\psi_l/\psi_k$ is actually a nonconstant polynomial. Thus we have shown the existence of a polynomial invariant.

\rightline{$\diamond$}
\medskip
The next result shows that (1.4) also holds for real Lie algebras with compact semisimple part.
\medskip
\noindent{\bf(1.5) Proposition.}
\smallskip
\noindent Let ${\bf K}={\bf R}$, let ${\cal M}$ such that $I_0({\cal M})\not={\bf R}$, and ${\cal L}$ the Lie algebra of a compact linear Lie group such that $\left[{\cal L},\,{\cal M}\right]\subseteq {\cal M}$. Then $I_0({\cal M}+{\cal L})$ is nontrivial.
\smallskip
\noindent{\it Proof.} Let $G$ be the compact and connected linear group with Lie algebra ${\cal L}$. Then $T{\cal M}T^{-1}={\cal M}$ for all $T\in G$, and $\psi\circ T^{-1}\in I_0({\cal M})$ for all $\psi\in I_0({\cal M})$ and $T\in G$.

\noindent Now let $\phi\in I_0({\cal M})$ be homogeneous and nonconstant, with $\phi(x)\geq 0$ for all $x$. (Take the square, if necessary.) The standard trick of setting
$$\hat\phi(x):=\int_G \phi(T^{-1}x) \,{\rm d}\sigma_G(T)$$
(integration with respect to Haar measure) produces an ${\cal M}$-invariant polynomial that is also $G$-invariant. Due to $\phi(0)=0$ and positivity, $\hat\phi$ is not constant.

\rightline{$\diamond$}
\medskip
A way to attempt a general proof of the converse to (1.1) - provided that it always holds - might use the existence of a Levi decomposition (semisimple subalgebra + radical). If the semisimple part is compact then (1.5) is applicable. However, extending or modifying the argument for arbitrary semisimple (or reductive) algebras seems to be nontrivial.

To finish this section we point out that ${\cal C}({\cal M})$ is not only a Lie algebra, but that actually the ``individual constituents'' of the Lie bracket of two elements of ${\cal C}({\cal M})$ are themselves in ${\cal C}({\cal M})$. This property depends essentially on the linearity of ${\cal M}$.
\medskip
\noindent{\bf (1.6) Proposition.}
\smallskip
\noindent If $p(x)$, $q(x)$ are in ${\cal C}({\cal M})$ then also $Dp(x)q(x)\in {\cal C}({\cal M})$.
\smallskip
\noindent{\it Proof.} Let $B\in{\cal M}$. Then $Dp(x)Bx=Bp(x)$, whence
$$D^2p(x)(Bx, \,y)+Dp(x)By=B\,Dp(x)y$$
by differentiation. Substitute $q$ and use $Bq(x)=Dq(x)Bx$ to obtain
$$D^2p(x)(Bx, \,q(x))+Dp(x)Dq(x)Bx=B\,Dp(x)q(x),$$
as was to be shown.

\rightline{$\diamond$}
\medskip
For the finite dimensional centralizer case this forces a certain ``nilpotency property'' on nonlinear centralizer elements.
\medskip
\noindent{\bf (1.7) Corollary.}
\smallskip
\noindent Let ${\cal C}({\cal M})$ be finite dimensional, and $d$ the maximal degree of elements in ${\cal C}({\cal M})$. If $d>1$, and $p$, $q$ are homogeneous elements of ${\cal C}({\cal M})$ such that ${\rm deg}\,p +{\rm deg}\,q>d+1$, then $Dp(x)q(x)=0$.

\rightline{$\diamond$}

\noindent 
\bigskip  
\noindent{\mittel 2. Some examples and applications.}
\smallskip
\noindent We first present some examples concerning Poincar\'e-Dulac normal forms.
\medskip
\noindent{\bf(2.1) Example.} (One-parameter symmetry group with ``simple resonance''.)
\smallskip
\noindent Consider the one-dimensional Lie algebra spanned by $A={\rm diag}\,(\sigma_1,\ldots, \sigma_n)$. Assume that there are integers $d_1,\ldots,d_n>0$ with $d_1\sigma_1+\ldots+d_n\sigma_n=0$, and that whenever $m_1\sigma_1+\ldots+m_n\sigma_n=0$ with integers $m_i\geq 0$ then $(m_1,\ldots,m_n)\in {\bf Z}\cdot(d_1,\ldots, d_n)$. Define $\psi(x):= x_1^{d_1}\cdots x_n^{d_n}$. The above and the familiar symmetry conditions enforce that a homogeneous polynomial vector field $p$ lies in ${\cal C}({\bf K} A)$ if and only if $p=\psi^k\cdot C$, with some integer $k\geq 0$ and a diagonal matrix $C$. To see this, use the fact that the polynomials in ${\cal C}({\bf K} A)$ are precisely the linear combinations of those monomials $x_1^{r_1}\cdots x_n^{r_n}\cdot e_\ell$ (with $e_1,\ldots,e_n$ the standard basis) which satisfy $r_1\sigma_1+\ldots+r_n\sigma_n-\sigma_\ell=0$.

\noindent{\it Suppose that the analytic vector field
$$f=B+\sum_{j\geq 1}f_j$$
(with each $f_j$ homogeneous of degree $j+1$) is such that the eigenvalues of $B$ satisfy Bruno's Condition $\omega$ (see [2]), and that $f$ is centralized by ${\bf K} A$. If the abelian algebra ${\cal M}={\bf K}A+{\bf K}B$ has no nontrivial polynomial integral then $f$ admits a convergent transformation to normal form, and this normal form is linear. }

\noindent To verify this, recall that there is a formal transformation to normal form $\hat f$ which respects the infinitesimal symmetry $A$, whence
$$\hat f = B+\sum_{k\geq 1} \psi^k\cdot C_k.$$
Since $X_B(\psi)\not= 0$, $\left[B,\,\hat f\right] =0$ is only possible if all $C_k=0$, and we see that the formal normal form $\hat f = B$ is linear, whence Bruno's Condition A is satisfied. Existence of a convergent transformation now follows from Bruno [2] (see also Pliss [10]). We note that Cicogna [5] uses a similar argument. Here we have a case where ${\cal C}({\cal M})$ is not only finite dimensional but contains only linear vector fields.

\rightline{$\diamond$}
\medskip
\noindent{\bf (2.2) Example.}
\smallskip
\noindent Consider the one-dimensional subalgebra of $gl(3,\,{\bf R})$ which is spanned by 
$$A={\rm diag}\,(-1,2,3).$$
Moreover, let the analytic vector field
$$f(x)={\rm diag}\,(-1,-1,0)\cdot x +\ldots$$
be centralized by $A$. 
Since the powers $x_3$ are the only homogeneous polynomial invariants of $B:={\rm diag}\,(-1,-1,0)$, $B$ and $A$ have no common invariant, whence any normal form $\hat f$ of $f$ has a finite Taylor expansion. A direct computation shows
$$\hat f(x) = Bx+\theta\cdot\pmatrix{0\cr x_1x_3 \cr 0 \cr}.$$
This situation is topologically nontrivial, since $\hat f$ has a non-isolated (stable) stationary point.

\noindent While there seems to be no general result guaranteeing the existence of a convergent transformation from $f$ to $\hat f$, the formal information on $\hat f$ is sufficient to guarantee a convergent transformation of $f$ to normal form on an invariant manifold (abbreviated NFIM; see Bibikov [1], Ch.~I, Theorem 3.2). This means that $f$ is conjugate to a vector field
$$f^*(x) = Bx+\pmatrix{x_1\phi_{11}(x)+x_2\phi_{12}(x)\cr
                       x_1\phi_{21}(x)+x_2\phi_{22}(x)\cr
                       \phi_3(x_1,\,x_2)\cr},$$
where the initial terms of $\phi_3$ are of order $>1$, and the $\phi_{kl}$ have zero constant terms (see Bibikov [1], Ch.~I, Definition 3.1). Therefore 0 is a non-isolated stationary point of $f^*$ (as well as $f$), and one directly verifies the existence of a first integral of type $\rho(x)=x_3 +\rho^*(x)$, with $\rho^*$ of order $>1$. 
This ensures stability of the stationary point of $f$.

\noindent Note that the group invariance property was critical in allowing this conclusion.

\rightline{$\diamond$}
\medskip
Now we turn to properties of nonlinear (and not necessarily autonomous) equations that are associated with finite dimensional centralizers of Lie algebras of linear Lie groups. Such equations are of some theoretical interest since they admit symmetries but there is no nontrivial notion of a reduced phase space (in other words, such equations cannot be reduced to smaller dimension by invariants). Therefore a different type of reduction is required. We will show that there is  a direct elementary approach.
\medskip
\noindent{\bf (2.3) Proposition.}
\smallskip
\noindent Let ${\cal M}$ be such that ${\cal C}({\cal M})$ is finite dimensional, and let $p_1,\ldots,p_r\in {\cal C}({\cal M})$ be homogeneous, with degrees $>1$. Moreover, let $\alpha_1,\ldots,\alpha_r\in{\bf K}$, and let $\sigma_1,\ldots,\sigma_r$ be polynomials. Then every solution of
$$\dot x =\sum_{j=1}^r \sigma_j(t){\rm e}^{\alpha_j t}p_j(x)$$
is elementary, i.e., every initial value problem has a solution of the form $\sum \rho_l(t){\rm e}^{\alpha_l t}u_l$, with polynomials $\rho_l$, and the $u_l$ are certain elements of ${\cal C}({\cal M})$, evaluated at the initial value. In the autonomous case (constant $\sigma_j$ and $\alpha_j=0$) the solution is polynomial.
\smallskip
\noindent{\it Proof (sketch).} The solution for a fixed initial value $y$ (at $t=0$, say) has a power series expansion in $t$, and its coefficients can be determined by iteratively computing derivatives of higher order. Computing these derivatives will yield terms of type $Dp_j(y)p_k(y)$, and iterates of such terms, which by (1.6) themselves correspond to elements of ${\cal C}({\cal M})$. Due to the nilpotency property of (1.7), there is only a finite number of such iterates. It is elementary to see that summation produces only terms of the form $\rho_l(t){\rm e}^{\alpha_l t}$.

\rightline{$\diamond$}
\medskip
There is a different perspective from which this result may be viewed. Equations of this type can be ``linearized by adding variables'', more precisely, by augmenting the coordinate vector $x$ with a basis of the space of vector fields of order $>1$ in $ {\cal C}({\cal M})$, and considering the larger system that results. Then (1.6) and the proof of (2.3) show that this procedure leads to a finite dimensional linear system. Moreover, suitable ordering will yield a system that is strictly lower triangular. (The procedure is described in detail, albeit in the context of normal forms, in [8].)
\medskip
\noindent {\bf (2.4) Corollary.}
\smallskip
\noindent Let the hypotheses on ${\cal M}$ be as above. Then every solution of 
$$\dot x = q(x),\quad{\rm with}\quad q\in{\cal C}({\cal M}),\,\,q(0)=0$$
is elementary.
\smallskip
\noindent{\it Proof.} Write $q=B+\tilde q$, with $B$ linear and $\tilde q$ collecting the nonlinear homogeneous terms. Thus we have
$$\dot x = Bx+\tilde q(x).$$
Using an argument by Chen [4], we let $T(t)$ solve the matrix equation $\dot T=BT$, $T(0)=E$, and set $x=Tz$. This yields
$$\dot z = T(t)^{-1}\tilde q(T(t)z),$$
and the formula for adjoint action
$$T(t)^{-1}\tilde q(T(t)z)=\exp(t\,{\rm ad}\,B)q(z)$$
implies that $T^{-1}\tilde q(Tz)$ can be rewritten as required in (2.3). (This involves a standard linear algebra argument: One shows that the polynomials corresponding to a Jordan canonical form of ${\rm ad}\,B$ are all contained in ${\cal C}({\cal M})$.) The result follows.

\rightline{$\diamond$}
\medskip
Special cases of this are known; see for instance [16], Cor.~2.5, on normal forms when the linear part admits no polynomial integral.
\bigskip
\bigskip
\noindent{\it Acknowledgements.} This paper was written during the authors' visit at the University of Cagliari. We wish to express our thanks to the Mathematics Department of this University, in particular to Prof. T.~Gramchev, for their hospitality. The work of G.G.~was supported by Fondazione CARIPLO.
\bigskip
\bigskip
\bigskip
\noindent{\mittel References.}
\medskip
\noindent[1] Yu.N. Bibikov: {\it Local theory of nonlinear analytic ordinary differential equations.} Lecture Notes in Mathematics {\bf 702}, Springer-Verlag, Berlin (1979).
\smallskip
\noindent[2] A.D.~Bruno: {\it Local methods in nonlinear differential equations.} Springer-Verlag, Berlin (1989).
\smallskip
\noindent[3] J.F.~Cari\~nena, J.~Grabowski, G.~Marmo: {\it Lie-Scheffers systems: A geometric approach.} Bibliopolis, Napoli (2000).
\smallskip
\noindent[4] K.-T.~Chen: {\it Decomposition of differential equations.} Math. Ann. {\bf 146}, 263 - 278 (1962).
\smallskip
\noindent[5] G.~Cicogna: {\it On the convergence of normalizing transformations in the presence of symmetries.} J. Math. Anal. Appl. {\bf 199}, 243 - 255 (1996).
\smallskip
\noindent[6] G.~Cicogna, S.~Walcher: {\it Convergence of normal form transformations: The role of symmetries.} Acta Appl. Math. (to appear)
\smallskip
\noindent[7] G.~Gaeta: {\it Nonlinear symmetries and nonlinear equations.} Kluwer, Dordrecht (1994).
\smallskip
\noindent[8] G.~Gaeta: {\it Resonant normal forms as constrained linear systems.} Preprint, 16 pp. math-ph/0106017 (2001).
\smallskip
\noindent[9] S.~Lie, G.~Scheffers: {\it Vorlesungen \"uber continuierliche Grupen.} Teubner, Leipzig (1893).
\smallskip
\noindent[10] V.A.~Pliss: {\it On the reduction of an analytic system of differential equations to linear form.} Differential Equations {\bf 1}, 111-118 (1965).
\smallskip
\noindent[11] G.~Sartori: {\it Orbit spaces of algebraic reductive groups as phase spaces generated by spontaneous symmetry breaking and/or supersymmetry breaking.} Acta Appl. Math. (to appear).
\smallskip
\noindent[12] S.~Shnider, P.~Winternitz: {\it Nonlinear equations with superposition principles and the theory of transitive primitive Lie algebras.} Lett. Math. Phys. {\bf 8}(1), 69 - 78 (1984).
\smallskip
\noindent[13] S.~Shnider, P.~Winternitz: {\it Classification of nonlinear ordinary differential equations with superposition principles.} J. Math. Phys. {\bf 25}(11), 3155 - 3165 (1984).
\smallskip
\noindent[14] G.W.~Schwarz: {\it The topology of algebraic quotients.} In: {\it Topological methods in algebraic transformation groups (New Brunswick, NJ, 1988).} Progr. Math. {\bf 80}, 135 - 151 (1989). 
\smallskip
\noindent[15] S.~Walcher: {\it \"Uber polynomiale, insbesondere Riccatische, Differentialgleichungen mit Fundamentall\"osungen.} Math. Ann. {\bf 275}, 269 - 280 (1986).
\smallskip
\noindent[16] S.~Walcher: {\it On differential equations in normal form.} Math. Ann. {\bf 291}, 293 - 314 (1991).
\smallskip
\noindent[17] S.~Walcher: {\it Multi-parameter symmetries of first-order ordinary differential equations.} J.~Lie Theory {\bf 9}, 249 - 269 (1999).
\smallskip
\noindent[18] P.~Winternitz: {\it Nonlinear action of Lie groups and superposition principles for nonlinear differential equations.} Proceedings of the Tenth International Colloquium on Group-Theoretical Methods in Physics (Canterbury, 1981). Phys. A {\bf 114}, 105 - 113 (1982).
\smallskip
\noindent[19] P.~Winternitz: {\it Comments on superposition rules for nonlinear coupled first-order differential equations.} J. Math. Phys. {\bf 25} (7), 2149 - 2150 (1984).

\bye